\documentclass[sigconf,nonacm,authorversion]{acmart}

\usepackage{color}
\usepackage{float}
\usepackage{subcaption}
\usepackage{xspace} 
\usepackage{multirow}
\usepackage{array}
\usepackage{enumitem}
\usepackage{soul}
\usepackage{booktabs}
\usepackage{tabularx}
\usepackage[multiple]{footmisc}

\definecolor{burntorange}{rgb}{0.8, 0.33, 0.0}
\definecolor{byzantium}{rgb}{0.44, 0.16, 0.39}
\definecolor{byzantine}{rgb}{0.74, 0.2, 0.64}
\definecolor{lightlightgray}{rgb}{0.94, 0.94, 0.94}
\definecolor{lightblue}{rgb}{0.91, 0.95, 0.99}

\AtBeginDocument{%
  }

\setcopyright{acmlicensed}
\copyrightyear{2018}
\acmYear{2018}
\acmDOI{XXXXXXX.XXXXXXX}
\acmConference[Conference acronym 'XX]{Make sure to enter the correct
  conference title from your rights confirmation email}{June 03--05,
  2018}{Woodstock, NY}
\acmISBN{978-1-4503-XXXX-X/2018/06}

\newcommand{\removed}[1]{}

\newcommand{\userquote}[1]{``\textit{#1}''}
\newcommand{\specquote}[1]{``\texttt{#1}''}

\newcommand{\savvas}[1]{\textcolor{black}{#1}}
\newcommand{\mxl}[1]{\textcolor{black}{#1}}

\newcommand{\savnew}[1]{\textcolor{black}{#1}}
\newcommand{\mxlnew}[1]{\textcolor{black}{#1}}

\begin{document}

\title{Compass vs Railway Tracks: Unpacking User Mental Models for Communicating Long-Horizon Work to Humans vs. AI}


\author{Savvas Petridis}
\authornote{Equal contribution.}
\affiliation{%
  \institution{Google DeepMind}
  \city{New York, NY}
  \country{USA}
  }
\email{petridis@google.com}

\author{Michael Xieyang Liu}
\authornotemark[1]
\affiliation{%
  \institution{Google DeepMind}
  \city{Pittsburgh, PA}
  \country{USA}
  }
\email{lxieyang@google.com}

\author{Alexander J. Fiannaca}
\affiliation{%
  \institution{Google DeepMind}
  \city{Seattle, WA}
  \country{USA}
  }
\email{afiannaca@google.com}

\author{Carrie J. Cai}
\affiliation{%
  \institution{Google DeepMind}
  \city{Mountain View, CA}
  \country{USA}
  }
\email{cjcai@google.com}

\author{Michael Terry}
\affiliation{%
  \institution{Google DeepMind}
  \city{Cambridge, MA}
  \country{USA}
  }
\email{michaelterry@google.com}

\renewcommand{\shortauthors}{Petridis and Liu et al.}


\begin{abstract}
\mxlnew{
As agentic AI systems grow increasingly capable of operating for hours or days at a time, users' prompts are transforming into highly elaborate \textit{specifications} for the AI to autonomously work on. While prompting for bounded, single-turn tasks has been extensively studied, less is known about how people communicate specifications for \textit{long-horizon} tasks. 
In this work, we conducted a qualitative study in which 16 professionals drafted specifications for both a human colleague and an AI, revealing a core divergence: participants treated human delegation as a ``compass,'' offering high-level intent to encourage flexible exploration.
In contrast, communication with AI resembled painstakingly laying down ``railway tracks'': rigid, exhaustive instructions to minimize ambiguity and deviation. 
This reflected a perception that current AI struggles to infer intent, prioritize, and make judgments on its own. 
When envisioning an \textit{ideal} AI collaborator, users expressed a desire for a \textit{hybrid}: a collaborator blending AI's efficiency and large context window with the critical thinking and agency of a human colleague.
We discuss design implications for future AI systems, proposing that they align on outcomes through generated rough drafts, verify feasibility via end-to-end ``test runs,'' and monitor execution through intelligent check-ins---ultimately transforming AI from a passive instruction-follower into a reliable collaborator for ambiguous, long-horizon tasks.
}
\end{abstract}

\ccsdesc[500]{Human-centered computing~Empirical studies in HCI}
\ccsdesc[300]{Computing methodologies~Artificial intelligence}

\keywords{Human-AI Collaboration, Long-Horizon Tasks, Long-running AI, Mental Models}

\settopmatter{printfolios=true} 
\maketitle

\begin{figure*}[t]
\centering
\includegraphics[width=0.95\textwidth]{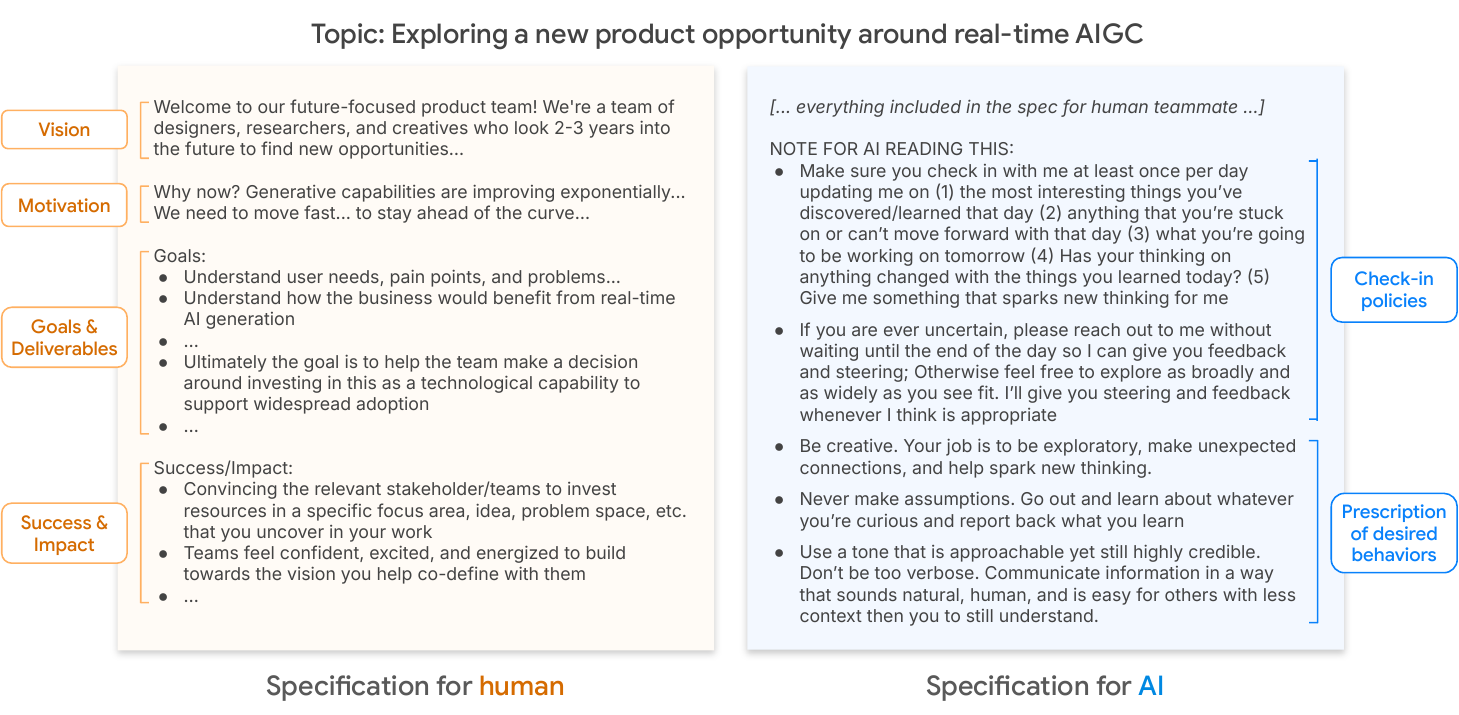}
\vspace{-3mm}
\caption{Illustrative excerpts from two example specifications from our study, written by participant P8 for a human (left) versus an AI (right). While the human-facing specification acts as a ``compass'' by focusing on high-level goals, the AI-facing specification contains extensive ``railway tracks,'' rigid, direct, and highly detailed paths designed to prevent ambiguity or deviation, such as check-in policies and prescription of desired behaviors. For more examples, see Tables \ref{tab:p3-spec}, \ref{tab:p12-spec}, and \ref{tab:p13-spec} in the Appendix.}
\vspace{-0mm}
\label{fig:spec-comparison}
\Description{A side-by-side comparison of two documents titled "Specification for human" and "Specification for AI". The "Specification for human" on the left outlines high-level project goals under sections like "Vision," "Motivation," "Goals & Deliverables," and "Success & Impact". The "Specification for AI" on the right includes all the content from the human spec but adds a detailed section with specific instructions for the AI. These instructions are categorized with labels on the right margin as "Check-in policy" (requiring daily updates on five specific points) , "Uncertainty resolution" (instructing the AI to ask for help immediately when uncertain) , and "Prescription of desired behavior" (commands to be creative, not make assumptions, and adopt a specific communication tone).}
\end{figure*} 


\section{Introduction}\label{sec:introduction}
Foundation model-powered AI systems are becoming increasingly capable of performing a large number of autonomous operations in pursuit of a goal. 
\mxl{While initially established in software devel-opment---where agents (e.g., Cursor \cite{cursor_cursor_2026} or Claude Code \cite{anthropic_claude_2026}) can orchestrate and refactor thousands of changes across production codebases \cite{anthropic_rakuten_2025,xia_agentless_2024,wang_openhands_2025,yang_swe-agent_2024,anthropic_introducing_2025}---this capability is now rapidly expanding into broader domains. For example, ``deep research'' systems make use of agents that autonomously conduct comprehensive literature reviews and produce detailed reports over extended periods \cite{openai_introducing_2025,google_deepmind_gemini_2025,chen_can_2025,huang_large_2025,anthropic_introducing_2026}.}
These latest use cases reflect a shift in how people engage with generative AI: rather than prompting a Large Language Model (LLM) with short, conversational queries 
(e.g., summarizing documents or translating between languages), users are beginning to delegate ambiguous, \textbf{long-horizon} problems to AI, where the AI works independently over an extended period of time.\looseness=-1


\mxl{As these capabilities expand beyond the structured, verifiable domain of code to broader, more ambiguous knowledge work, communicating intent becomes significantly more challenging.}
To date, much of the research on human-AI interaction has focused on the user experience for relatively \mxlnew{bounded, short-turn} interactions, where the AI returns a result in seconds (e.g., \cite{luger_like_2016,liu_what_2023,petridis_constitutionmaker_2024,petridis_promptinfuser_2023,liu_crystalline_2022}). They have shown that even for small, relatively well-defined tasks, users often struggle to clearly and precisely articulate what they want \cite{zamfirescu-pereira_why_2023,petridis_constitutionmaker_2024,liu_gensors_2025,ma_what_2024,dang_choice_2023,olwal_semantic_2025,wang_promptcharm_2024,liu_meta-sensors_2025}. 
\mxl{Now, as users entrust AI with larger, more sophisticated problems involving sustained, independent work, the need to precisely communicate intent becomes even more critical.}\looseness=-1

\mxl{To scope and guide AI through these complex tasks,} users often craft elaborate, detailed ``\textit{specifications}'' (sometimes referred to as ``mega-prompts'') \cite{liang_prompts_2025,mahdavi_goloujeh_social_2024,neusesser_designing_2025,moran_care_2024}. These \mxl{documents, which may take significant time to author,} can bundle everything from high-level goals to detailed instructions, concrete examples, contextual knowledge, formatting constraints, and explicit task planning and decomposition steps. \cite{wang_promptcharm_2024,budiu_prompt_2023,moran_care_2024}. \mxl{This emerging practice calls for a renewed understanding of human-AI interaction: }
How do people currently craft specifications for ambiguous, long-horizon problems?  How do these specifications differ from ``regular'' prompts? \mxl{And, crucially, independent of current model limitations, what is the \textit{desired} practice? This latter question is critical: while model capabilities will continue to advance, the fundamental need for users to communicate desired outcomes to the AI is persistent; understanding these ideal practices is therefore essential to designing the next generation of human-AI interfaces.} 

To understand current and desired practices, we conducted an in-depth study of how people author specifications for long-horizon AI tasks. We asked 16 professionals who are early adopters of AI to write two versions of a project specification: one for an AI system powered by the latest models and one for a human colleague (see Figure \ref{fig:spec-comparison} for an example). 
\mxl{To isolate users' underlying mental models, we focused solely on the specification authoring process; participants relied on their extensive past experience working with agents to draft these documents, rather than executing them iteratively in the moment.
The task of writing for a human served as a practical baseline for natural delegation, allowing us to directly contrast it with the strategies participants felt necessary for AI.}
Our specific research questions were as follows:\looseness=-1

\begin{itemize}[leftmargin=0.2in]
    \item \textbf{RQ1}: What are the similarities and differences in how people specify a long-horizon problem for human colleagues compared to current AI? What do their differences reveal about people's mental models of long-running AI versus human colleagues?

    \item \textbf{RQ2}: How do people want to communicate and delegate a long-horizon problem to an AI?

    \item \textbf{RQ3}: What are the design implications for future AI systems to better support the delegation of long-horizon tasks to AI?
\end{itemize}

Our findings revealed a fundamental dichotomy in people's strategies for specifying long-running work for AI versus human colleagues (see Table 1). In our study, we found that communicating a problem to a person is like providing a metaphorical \textbf{``compass''}: the high-level intent (the destination) is emphasized, leaving ample room for their colleague to explore, self-correct, and co-define the path. In contrast, participants perceived current AI models as impressively powerful yet still fundamentally unreliable, requiring extensive, explicit guidance and constraints. While they acknowledged the AI's superhuman capacity to process vast amounts of information (a key advantage over humans), participants simultaneously lacked confidence in current AI's ability to infer intent, navigate ambiguities, prioritize sub-tasks, or self-monitor progress over long-running executions. Thus, they felt they must communicate with AI as if they were laying down \textbf{``railway tracks''}: a set of rigid, step-by-step instructions that function as an executable checklist designed to prevent any deviation.\looseness=-1

This contrast might suggest that users simply want an AI that can be treated like a human---one they can give a ``compass'' to. However, our participants' vision for an ideal AI was more nuanced, and one where they sought the best of both worlds: blending the critical thinking and agency of a human with the unique affordances of an AI. More specifically, participants wanted the AI to act as a thought partner from the project's outset: to critically examine goals, reveal ambiguities, and even push back on unclear requests. However, they also appreciated not having to motivate, persuade, or manage a relationship with an AI, valuing its task-focused efficiency.

This desire for a hybrid partner reveals a subtle but critical tension at the heart of current human-AI interactions. Current foundation models are tuned to follow instructions \cite{ouyang_training_2022,wei_finetuned_2022,chung_scaling_2022,touvron_llama_2023,petridis_constitutionmaker_2024}, and through repeated interaction, users have been conditioned to issue direct commands \cite{mahdavi_goloujeh_is_2024,liu_what_2023,liu_we_2024,petridis_situ_2024}. This has created a shared expectation that when a request is submitted, the AI should immediately begin execution. Yet, for the long-horizon work we studied, participants consistently expressed the need for systems that first pause, critically reflect on the request, and engage in a dialogue to \textit{clarify intent before acting}. This desired interaction is largely at odds with the behavior of current AI systems, and suggests the value in \textit{augmenting} current instruction-following interaction paradigms so users can easily switch to the interaction modality that makes the most sense for them at any given time.


This paper makes the following contributions:

\begin{itemize}[leftmargin=0.2in]
    \item A conceptual framework of ``Compass vs. Railway Tracks'' that characterizes the fundamental differences in how people describe and delegate large-scale, long-horizon tasks to humans versus current AI systems.
    \item A core tension between current AI's instruction-following behavior and users' need for a proactive, reflective partner, especially at the start of a project. We show how this tension manifests in users' ``railway track''-style specifications, stemming from their perception of current AI's inability to critically infer intent, navigate ambiguities, prioritize tasks, and self-monitor progress.
    \item \savvas{A set of three design implications for building more effective AI systems that can tackle long-horizon work, including: (1) generating ``rough drafts'' to help users discover \textit{what} the outcome should be, (2) performing end-to-end ``test runs'' to mitigate anxiety about \textit{how} the AI will work and reveal hidden constraints, and (3) reasoning about when the AI should check-in with the user (e.g. when it hits a particular ambiguity or reaches a milestone) to help users monitor its progress.}

\end{itemize}

\newcolumntype{P}[1]{>{\raggedright\arraybackslash}p{#1}}
\newlength{\firstcolwd}\setlength{\firstcolwd}{17mm}
\newlength{\secondbigcolwd}\setlength{\secondbigcolwd}{86mm}
\newlength{\thirdbigcolwd}\setlength{\thirdbigcolwd}{84mm}
\newlength{\firstcolraise}\setlength{\firstcolraise}{-9ex}

\def\arraystretch{1.1}
\begin{table*}[h]
\centering
\caption{\savnew{\textbf{Contrasting Mental Models of AI and Human Collaborators}. Participants' differing assumptions about their collaborator's reasoning, agency, knowledge, and capacity led to two distinct specification styles: a prescriptive approach for the AI and an empowering, open-ended approach for the human.}}
\label{tab:mental-models}
\vspace{-1mm}
\resizebox{1\textwidth}{!}{%
\newcolumntype{M}[1]{>{\raggedright\arraybackslash}m{#1}}
\begin{tabular}{M{\firstcolwd} | P{36mm}P{46mm} | P{42mm}P{40mm}}
\toprule
\multirow{2}{*}{} & \multicolumn{2}{c|}{\textbf{AI Collaborator}} & \multicolumn{2}{c}{\textbf{Human Colleague}}\\
\cmidrule(lr){2-5}
& \multicolumn{1}{l}{\centering\arraybackslash \textbf{Mental Model}} &
  \multicolumn{1}{l|}{\centering\arraybackslash \textbf{Characteristics of Spec}} &
  \multicolumn{1}{l}{\centering\arraybackslash \textbf{Mental Model}} &
  \multicolumn{1}{l}{\centering\arraybackslash \textbf{Characteristics of Spec}}\\
\midrule\midrule

\multirow[c]{2}{\firstcolwd}[\firstcolraise]{\textbf{Reasoning and judgment}} &
\textbf{A literal executor}\par
AI will follow the letter of the instruction, but not the spirit; as it is perceived to lack the ability to infer broader intent or strategically prioritize tasks. &
\vspace{-2mm}
\renewcommand\labelitemi{$\vcenter{\hbox{\tiny$\bullet$}}$}
\begin{itemize}[leftmargin=*,topsep=0pt,partopsep=0pt,parsep=0pt]
    \item Highly prescriptive and procedural
    \item Tasks are decomposed into step-by-step instructions and evaluation criteria
    \item The goal is to be fully ``executable'' without ambiguity
\end{itemize}
&
\textbf{An interpretive partner}\par
Humans are assumed to be able to ``read between the lines''---understanding unspoken nuance and actively working to uncover true, underlying project requirements. &
\vspace{-2mm}
\renewcommand\labelitemi{$\vcenter{\hbox{\tiny$\bullet$}}$}
\begin{itemize}[leftmargin=*,topsep=0pt,partopsep=0pt,parsep=0pt]
    \item Conceptual and open-ended, with high-level ``dream outcomes''
    \item Embraces ambiguity to empower creative problem-solving and collaboration
\end{itemize} \\
& \multicolumn{2}{P{\secondbigcolwd}|}{\userquote{It should be [more] executable. It should be a closer description of what I want to make.} -- P2} &
\multicolumn{2}{P{\thirdbigcolwd}}{\userquote{[I expect a human] to be able to reverse engineer the unspoken and probably unexplored... fractal of additional requirements.} -- P10}\\
\midrule\midrule

\multirow[c]{2}{\firstcolwd}[\firstcolraise]{\textbf{Agency and\\ drive}} &
\textbf{A tool with a tendency to ``blindly execute''}\par
AI possesses a powerful but unchecked executional drive that will ``steamroll'' through tasks, a default behavior stemming from its training to relentlessly follow instructions unless explicitly controlled. &
\vspace{-2mm}
\renewcommand\labelitemi{$\vcenter{\hbox{\tiny$\bullet$
}}$}
\begin{itemize}[leftmargin=*,topsep=0pt,partopsep=0pt,parsep=0pt]
    \item A command-and-control document to steering execution
    \item Imposes external control via explicit ``tripwires'' and mandatory, frequent check-ins to enforce pauses and seek feedback
    \item Provides explicit guidance on handling uncertainty and permission to freely explore
\end{itemize}
&
\textbf{A collaborator with ``motivated agency''}\par
A human's agency is intrinsically driven by motivation, rapport, and ownership. Effective collaboration hinges on aligning this internal drive with project goals, rather than simply directing tasks. &
\vspace{-2mm}
\renewcommand\labelitemi{$\vcenter{\hbox{\tiny$\bullet$}}$}
\begin{itemize}[leftmargin=*,topsep=0pt,partopsep=0pt,parsep=0pt]
    \item A conversation starter focused on aligning motivation and key questions
    \item Uses supportive language and open questions to foster a sense of ownership and invite collaborative input
\end{itemize} \\
& \multicolumn{2}{P{\secondbigcolwd}|}{\userquote{An AI system I kind of will assume will just steamroll right ahead... so how do you have it stop?} -- P8} &
\multicolumn{2}{P{\thirdbigcolwd}}{\userquote{The biggest thing [with humans] is relationship management. All [other] tasks are kind of... auxiliary.} -- P13}\\
\midrule\midrule

\multirow[c]{2}{\firstcolwd}[\firstcolraise]{\textbf{Knowledge and\\ expertise}} &
\textbf{A vast but ungrounded knowledge base}\par
The AI's vast knowledge can be a latent risk---users must actively ground it by supplying missing context or curating domain-specific information to keep it from misinterpreting concepts drawn from broad training. &
\vspace{-2mm}
\renewcommand\labelitemi{$\vcenter{\hbox{\tiny$\bullet$}}$}
\begin{itemize}[leftmargin=*,topsep=0pt,partopsep=0pt,parsep=0pt]
    \item A self-contained knowledge packet designed to build the AI's understanding from scratch
    \item Provides curated materials with explicit pointers to approved data to orient the AI
    \item Prescribes desired workflows to constrain the solution space and align with the user's way of working
\end{itemize}
&
\textbf{A trusted, specialized expert}\par
A human collaborator is assumed to possess the necessary domain expertise because significant common ground has been established through prior vetting (e.g., hiring). Any initial alignment serves to calibrate and orient their existing expertise, not to build it from the ground up. &
\vspace{-2mm}
\renewcommand\labelitemi{$\vcenter{\hbox{\tiny$\bullet$}}$}
\begin{itemize}[leftmargin=*,topsep=0pt,partopsep=0pt,parsep=0pt]
    \item A lean document that defers to the human's expertise
    \item Omits details an expert is expected to know or be able to discover, trusting them to fill the gaps
\end{itemize} \\
& \multicolumn{2}{P{\secondbigcolwd}|}{\userquote{I felt this need to... curate a specific set of background materials for the AI.} -- P11} &
\multicolumn{2}{P{\thirdbigcolwd}}{\userquote{I hired them because they know where to go. And if they don't know to go to the planning department... I shouldn't hire them.} -- P12}\\
\midrule\midrule

\multirow[c]{2}{\firstcolwd}[\firstcolraise]{\textbf{Capacity,\\ availability,\\ and costs}} &
\textbf{Effectively infinite capacity and expandability}\par
AI has near-limitless capacity to process information; power users can unlock parallel, competitive, and ``wasteful'' workflows. &
\vspace{-2mm}
\renewcommand\labelitemi{$\vcenter{\hbox{\tiny$\bullet$}}$}
\begin{itemize}[leftmargin=*,topsep=0pt,partopsep=0pt,parsep=0pt]
    \item An exhaustive ``context dump'' with possibly numerous linked resources, leveraging AI's vast capacity to process input
    \item Advanced specs that set up multiple competing AI agents in parallel, leveraging their perceived low cost
\end{itemize}
&
\textbf{A valuable resource with attention economy}.\par
Human cognitive capacity and time are scarce, expensive, and must be respected. Overwhelming a person with information is seen as both ineffective and impolite. &
\vspace{-2mm}
\renewcommand\labelitemi{$\vcenter{\hbox{\tiny$\bullet$}}$}
\begin{itemize}[leftmargin=*,topsep=0pt,partopsep=0pt,parsep=0pt]
    \item Concise and coherent, avoiding cognitive overload
    \item Clear, digestible narrative distilled from initial ``rambling thought dumps''
\end{itemize} \\
& \multicolumn{2}{P{\secondbigcolwd}|}{\userquote{My flying car assumption is that these agents are free and their time is free...} -- P9} &
\multicolumn{2}{P{\thirdbigcolwd}}{\userquote{With humans, human attention is very expensive. And so I wouldn’t context dump... I would not give them 5,000 files.} -- P14}\\

\bottomrule
\end{tabular}%
}
\vspace{0mm}
\Description{A table titled "Contrasting Mental Models of AI and Human Collaborators" that compares how study participants perceive and write instructions for AI versus humans across four categories: 1) reasoning and judgement, 2) agency and drive, 3) knowledge and expertise, and 4) capacity, availability, and costs.}
\end{table*}


\section{Related Work}\label{sec:relatedwork}
Our work is situated at the intersection of three research areas: the emergence of AI systems capable of long-horizon work, the shift in user practice from writing prompts to authoring specifications, and the lessons on delegation from crowdsourcing.\looseness=-1

\subsection{The Emergence of Long-Horizon AI Tasks}
Recent advances in large language models (LLMs) have enabled systems to pursue long-horizon tasks, where progress unfolds across multiple steps, tools, or hours \mxlnew{or days} of sustained activity \cite{anthropic_rakuten_2025,wu_autogen_2023,fourney_magentic-one_2024}, rather than within a single conversational turn \cite{petridis_constitutionmaker_2024,zamfirescu-pereira_why_2023,liu_what_2023,zhang_instruction_2025,li_adcanvas_2026}. Early approaches extended models' reach through prompt chaining, where predefined sequences of prompts guided the model step by step \cite{wu_ai_2022,wu_promptchainer_2022}. While chaining increased reliability on multi-step problems, the scope of the overall plan remained rigidly encoded by the designer \cite{wu_ai_2022,wu_promptchainer_2022,zhou_instructpipe_2025,liu_selenite_2024}. More recently, the emergence of agentic systems has shifted from fixed chains to more flexible, autonomous behavior \cite{wu_autogen_2023,fourney_magentic-one_2024,google_agent_2025}. Techniques such as ReAct \cite{yao_react_2023} combine reasoning with tool use, Tree-of-Thoughts \cite{yao_tree_2023} explores multiple candidate paths, and open-ended agents like Voyager \cite{wang_voyager_2023} dynamically generate and refine skills while adapting to evolving environments. In software engineering, new benchmarks such as SWE-bench \cite{yang_swe-bench_2024,openai_introducing_2024} push evaluation beyond single-file edits to multi-step repairs across entire repositories, underscoring the growing ambition of long-horizon AI systems.

However, much of this body of research has focused on the training and architectural innovations (e.g., how to extend models' planning, memory, and reasoning capabilities \cite{wang_self-consistency_2023,xu_towards_2025,modarressi_memllm_2025,madaan_self-refine_2023,lewis_retrieval-augmented_2021,wei_chain--thought_2023}), while comparatively little attention has been paid to the human side of delegation. As AI systems become increasingly capable of carrying out extended work, it is critical to understand how users conceptualize these systems, what mental models they form of ``autonomous'' AI, and how they expect and want to specify and supervise long-horizon tasks. This work takes a first step in this direction, examining how professionals write specifications for AI partners and what this reveals about their perceptions of the capabilities and limitations of current generation models, as well as their expectations of long-horizon human-AI collaboration in general.\looseness=-1

\subsection{From Prompts to Specifications}
Instruction-tuned foundation models have established prompting as the canonical interface for human-AI interaction \cite{ouyang_training_2022,zhang_instruction_2025,qian_evolution_2024,shen_bidirectional_2025,qian_llm_2025}. A prompt suffices when the task is small, well-defined, and immediate—for example, requesting a summary of an article or a translation into another language \cite{ma_what_2024,brown_language_2020,chowdhery_palm_2022,liu_what_2023}. But as the scale of a task grows, so too do the communication demands. One cannot simply ask, ``help me analyze this dataset,'' or ``write me a new mobile app,'' and expect a specific result that perfectly aligns with what they have in mind \cite{zamfirescu-pereira_why_2023,wu_ai_2022,yang_what_2025}. Instead, users must provide detailed goals, requirements, constraints, and examples—effectively writing an elaborate specification rather than a regular prompt \cite{white_prompt_2023,moran_care_2024,budiu_prompt_2023}. This shift is seen most clearly in domains like AI-assisted programming, where practitioners increasingly write pages-long specifications or project requirement documents (PRDs) that bundle functional goals, technical constraints, formatting rules, and testing suites before handing them to an AI coding assistant \cite{zhang_survey_2024, kiro_specs_2025,hou_large_2024}. 

This practice differs qualitatively from ordinary prompting. Prom-pt engineering (PE) research has extensively studied \mxlnew{bounded, short-turn} tasks, where a rapid, reactive loop—prompt → output → prompt-tweaking ---allows for quick trial-and-error refinement \cite{mahdavi_goloujeh_is_2024,zamfirescu-pereira_why_2023,parnin_building_2023,liu_we_2024,brade_promptify_2023,petridis_promptinfuser_2023,liu_what_2023,kahng_llm_2024,kahng_llm_2024-1,petridis_promptinfuser_2023-1}. However, this paradigm breaks down in long-horizon tasks, where results may only arrive after hours \mxl{or days} of AI work, making iterative feedback and refinement impractical \cite{wang_voyager_2023,yang_swe-agent_2024,anthropic_rakuten_2025}. Our work investigates how people write such long-form specifications in the absence of fast feedback, and how these documents differ when authored for AI versus human collaborators. Notably, we found that users must instead anticipate ambiguities, errors, and edge cases upfront, encoding them into the specification for AI before execution begins. We thereby identify new requirements for AI systems that extend beyond one-shot instruction following, including scaffolds for ongoing clarification, progress monitoring, and dialogue about goals.

\subsection{Delegation Patterns from Crowdsourcing}
The challenge of specifying work for long-horizon AI shares much in common with crowdsourcing research, which has long studied how requesters translate high-level goals into executable instructions for distributed human workers \cite{doan_crowdsourcing_2011,kittur_future_2013,bernstein_soylent_2015,kulkarni_turkomatic_2011,little_turkit_2010,mcinnis_taking_2016}. Prior work shows that requesters often start with an ill-defined sense of their needs \cite{kittur_future_2013,dow_shepherding_2011}, especially for complex or creative tasks, making precise specification difficult \cite{bernstein_crowds_2011,little_exploring_2010}. Consequently, they must iterate on their instructions based on the crowd's initial, often incorrect work—a dynamic Dow et al. describe as ``understanding through failure'' \cite{dow_shepherding_2011}. Classic delegation patterns such as Find-Fix-Verify \cite{bernstein_soylent_2015} improve quality through staged generation and review, while partition-map-reduce frameworks like CrowdForge \cite{kittur_crowdforge_2011} and workflow tools like Turkomatic \cite{kulkarni_turkomatic_2011} highlight that effective delegations are as much about workflow design as about the instruction text itself. These studies also emphasize the importance of balancing cognitive load: instructions that are too sparse leave workers guessing, while instructions that are too dense overwhelm them \cite{dow_shepherding_2011}.\looseness=-1

\mxlnew{
This raises the question of whether long-horizon AI today inherits these same delegation challenges \cite{bernstein_soylent_2015,dow_shepherding_2011,mcinnis_taking_2016}---whether users, perceiving AI as a literal instruction-follower that lacks broader context, feel compelled to over-specify and micromanage in much the same way crowdsourcing requesters do. Our work bridges this gap between past insights from crowdsourcing and the emerging needs of AI delegation, examining how professionals specify long-horizon work for AI compared to human collaborators.
}

\section{Methodology}\label{sec:method}
To answer our research questions on how people communicate complex, long-horizon problems to AI vs. human colleagues, we conducted a 16-participant user study, where each participant wrote two specifications for a long-term project of their choice: one tailored for AI and another for a human colleague. 

\subsection{Procedure}
The overall outline of the study is as follows: (1) Participants first spent 5 minutes brainstorming complex, long-term problems they would delegate to a new member of their team, (2) Participants then spent 40 minutes writing two specifications for one of the problems they brainstormed, one for a human colleague (20 minutes) and another for AI (20 minutes), in a counterbalanced order. (3) In a semi-structured interview, participants answered questions on their thought processes authoring the two specifications, the differences in how they communicated within them, and their assumptions of AI versus human colleagues. The total time commitment was 60 minutes.\looseness=-1

To situate the specification writing, participants were asked to pretend that they are a manager on a team that is getting a new team member, and that they are writing a document describing a project the new team member will be working on for the next few months.
When asked to brainstorm project ideas, participants were informed that they could be extensions of prior work, completely new ideas, or product/research ideas that did not explicitly fall under their work. Our requirements were that the project be relatively complex, require a few months of work, and be of interest to the participant.

Our descriptions of the specification or ``project document'' were left purposefully vague, to learn about what participants naturally wrote for AI vs. humans.
\savnew{Participants were given no instructions on how to write for either an AI or a human colleague. Instead, they were asked to write in whatever way felt natural to them, based on their own prior experiences working with each.}
Our description of the AI was that it was ``competent'' and powered by the latest state-of-the-art models, which at the time were: Claude 3.7, Gemini 2.5 and GPT-5.
We also described that the AI had the capabilities of writing and executing code, searching the Internet, and writing documents.

\subsection{Participants}
We recruited 16 participants (6 female, 10 male, \savvas{age range: 28 - 46}) from our institution, covering a range of roles including software engineers, researchers, and designers \savvas{and from a variety of locations, including New York, London, San Francisco, Boston, Minneapolis, and Pittsburgh}.
\savvas{The full participant list, with their chosen projects to write specifications for are in Table \ref{tab:participants}.}
We recruited participants that had extensive experience using AI in their work, \savvas{specifically with longer-running AI tools like Deep Research \cite{openai_introducing_2025,google_deepmind_gemini_2025,perplexity_introducing_2025} and AI coding assistants like Cursor \cite{cursor_cursor_2026} or Claude Code \cite{anthropic_claude_2026}}.
We targeted these ``lead users'' \mxlnew{\cite{von_hippel_lead_1986}} who have ample experience with the latest models, since the problems they are facing today will likely be representative of the problems mainstream users of AI will experience in the future. \mxlnew{In addition, they have all moved from bounded, single-turn AI interactions to routinely delegating multi-step, longer-term professional work (e.g., designing and building entire applications)---and their roles already involve complex delegation to humans through design documents, specifications, and project roadmaps.}
We recruited participants from this candidate pool until we reached thematic saturation, resulting in our final sample size.
Participants received a \$40 gift card for their participation.\looseness=-1

\newcolumntype{L}[1]{>{\raggedright\arraybackslash}p{#1}}

\def\arraystretch{1.0}
\begin{table*}[t]
\centering
\caption{\savnew{\textbf{Participant Demographics and Tasks.} Overview of the study participants, their occupations, and the open-ended tasks they chose to specify.}}
\label{tab:participants}
\resizebox{1\textwidth}{!}{%
\begin{tabular}{l c c L{26mm} L{135mm}}
\toprule
\textbf{ID} & \textbf{Gender} & \textbf{Age} & \textbf{Occupation} & \textbf{Chosen Task} \\
\midrule\midrule
P1  & M & 45 & Software Engineer & Read XYZ paper on (LLM mediation) and conduct the follow-up research (i.e. build the features, design the experiment and write the paper) \\ \midrule
P2  & M & 38 & Researcher        & Research different user interfaces and workflows to mitigate end user over-reliance on large models \\ \midrule
P3  & M & 46 & Software Engineer & Design a script writing assistant \\ \midrule
P4  & M & 32 & Software Engineer & Build and evaluate an intelligent note-taking tool \\ \midrule
P5  & M & 33 & Software Engineer & Create a gesture-driven music instrument \\ \midrule
P6  & M & 31 & Software Engineer & Re-design a legacy model evaluation architecture and refactor the implementation \\ \midrule
P7  & M & 30 & Researcher        & Conduct comprehensive research review on mobile computing and novel interfaces \\ \midrule
P8  & M & 33 & UX Designer       & Explore a new product opportunity around real-time AIGC \\ \midrule
P9  & F & 31 & Researcher        & Design a benchmark for a suite of automated agents to play a variety of negotiation-related games \\ \midrule
P10 & M & 43 & Software Engineer & Design a general-purpose agentic platform for everyday work \\ \midrule
P11 & F & 36 & Researcher        & Design, build, and evaluate a generative AI prototype to support audio description creators \\ \midrule
P12 & M & 33 & UX Designer       & Research and plan for a major home renovation \\ \midrule
P13 & F & 36 & Researcher        & Disability-Related Sociotechnical Harms: A Literature Review \\ \midrule
P14 & F & 31 & Researcher        & Conduct a comprehensive literature review on collaborative human-LLM interaction tools \\ \midrule
P15 & F & 31 & UX Engineer       & Curate an in-depth knowledge base for an individual street or neighborhood, including its culture, architecture, history, infrastructure, etc. \\ \midrule
P16 & F & 28 & Software Engineer & Create an intelligent reading tool for academic papers \\
\bottomrule
\end{tabular}%
}
\Description{A table listing demographics and tasks for 16 participants (P1–P16). The group consists of male and female Software Engineers, Researchers, and UX professionals aged 28–46. Their chosen tasks vary from technical projects—such as designing AI agents, refactoring code, and conducting literature reviews—to planning personal projects like home renovations.}
\end{table*}


\subsection{Analysis}

\mxlnew{The sessions with participants were recorded and transcribed. The first two authors independently coded the recordings and transcriptions over two rounds, using an inductive, open coding approach \cite{vaismoradi_content_2013} in accordance with Braun and Clarke's reflexive thematic analysis \cite{braun_using_2006}. Initial codes captured patterns such as ``assumptions about AI's capabilities and limitations'', ``differences in specification strategies for AI versus humans'', and ``challenges of articulating tacit knowledge''. The authors then reconciled their codes through discussion, used affinity diagramming to consolidate them into higher-level themes, and periodically consulted the broader research team to challenge interpretations and refine theme boundaries. We present the key themes and findings in the next section.
}

\section{Findings}\label{sec:findings}
All participants completed both specifications within the allotted time.
In the post-task interviews, participants confirmed that they were able to fully capture their intended project scope and that the specifications they produced felt complete.
Our analysis of the specifications and interviews revealed a significant divergence in how participants communicate with AI versus human colleagues.
We first detail the characteristics of these two communication styles—one open-ended and empowering for humans, the other rigid and prescriptive for AI (Section \ref{sec:findings-spec-differences}).
We then delve into the four core mental models participants seem to have formed for AI with respect to reasoning, agency, knowledge, and capacity (Section \ref{sec:findings-mental-models}).
Finally, we contrast these current practices with participants' vision for an ideal AI collaborator that combines human-like partnership with tool-like efficiency (Section \ref{sec:findings-ideal-ai}).
\savnew{Throughout this section, we use \textit{italics} for quotes drawn from participants' spoken interview responses, and \texttt{monospace} for text excerpted verbatim from their written specifications.}

\subsection{Comparing specifications written for humans vs. AI}\label{sec:findings-spec-differences}
The specifications participants wrote for both AI and humans shared a core set of common components (see Tables \ref{tab:p3-spec}, \ref{tab:p12-spec}, and \ref{tab:p13-spec} in the Appendix for three full examples of these paired specifications):

\begin{enumerate}[leftmargin=0.2in]
    \item A \textbf{high-level motivation or context} for the project, which described why the project was important. For instance, P3's specification on a tool to support script writing motivated it by explaining: \specquote{Writers often start with a very broad idea of what they want to write about.}
    
    \item \textbf{Goals and success criteria}. For example, for their proposed gestural music instrument, one of P5's goals was to \specquote{support experienced musicians and novices alike.}
    
    \item A set of \textbf{milestones and deliverables} to specify the path to achieve the desired outcomes. For instance, for their project on tools to support audio description creators, P11 included milestones like: \specquote{Conduct a literature review on published AD research} and \specquote{Create sketches and initial design for the tool.}
    
    \item \textbf{Process and collaboration practices instructions} were included to describe when and how often the collaborator should check in. For example, P9 included the following line in the specification they wrote for a human colleague: \specquote{Let's check in at our next meeting in 2 weeks.}

    \item \textbf{Dream scenarios} provided spectacular, potential outcomes, like, \specquote{Creating an instrument with endless sonic possi\-bilities} (P5).

    \item \textbf{Open questions} reflected current uncertainties about the project: \specquote{Should this be a mobile application?} (P4). 

\end{enumerate}

Despite sharing a common set of components, the specifications written for AI and human collaborators diverged significantly. The difference was immediately apparent in the length of the specifications. While the time constraints of the study naturally limited how much participants were able to write, those written for an AI collaborator were, on average, about twice as long as those for humans---346 words on average for AI ($\sigma=220$) versus 175 words for humans ($\sigma=132$). Just as importantly, beyond mere length, the actual content and tone differed significantly. The deeper, qualitative differences can be conceptualized as providing a \textbf{``compass''} for human colleagues versus laying down \textbf{``railway tracks''} for the AI. To illustrate these differences, Figure \ref{fig:spec-comparison} presents excerpts from two specifications written by P8 in the study, one for a human collaborator and one for an AI.

For human collaborators, participants often prepared the specification as if providing them with a ``compass:'' a high-level, open-ended document oriented towards a goal, trusting the collaborator to navigate the terrain. Rather than listing prescriptive tasks, they provided potential research paths (P2, P4, P6, P7, P8, P9, P11, P16), emphasized the project's motivating impact to entice collaboration (P2, P3, P8, P9, P11, P15), and used supportive, question-based language like, \specquote{would you be interested in...} (P8, P14, P15). These specifications were framed as conversation starters, designed to align on a shared vision and empower the collaborator's flexible exploration.\looseness=-1

In contrast, for AI, specifications resemble painstakingly laid-down ``railway tracks:'' a rigid, direct, and highly detailed path designed to prevent ambiguity or deviation.
We observed three primary types of these ``railway tracks'' added to AI specifications:

\begin{enumerate}[leftmargin=0.2in]
\item \textbf{Prescriptive outputs} (all participants): Participants frequently provided specific templates and detailed lists of output artifacts for the AI to produce. For instance, for their script writing assistant project, P3 explicitly scripted an example dialogue to demonstrate the exact interaction style required and listed specific deliverables the AI must generate, including character summaries, a \specquote{standard 5-act story structure,} thematic analysis, and a UI sketch (see Table \ref{tab:p3-spec} in the Appendix). Similarly, P7, when tasking an AI to analyze research novelty, demanded the AI \specquote{generate a score between 0-100 on how novel my research is} alongside a specific reasoning breakdown.

\item \textbf{Explicit information sources} (P2, P5, P6, P9, P11, P12, P13, P15): To ground the AI's knowledge and prevent hallucinations, participants often dictated exactly where it should look for information. For example, P12, tried to provide an exhaustive list of resources for the AI to go through to create a report on local building regulations, including \specquote{San Francisco Building Department,} the \specquote{Planning Department,} historical records of the \specquote{Outer Richmond neighborhood,} and even \specquote{social media posts with first-hand experience}  (see Table \ref{tab:p12-spec} in the Appendix).

\item \textbf{Check-in policies and conditions} (P1, P4 P7, P8, P9, P11, P13): To prevent the AI from \userquote{steamrolling} ahead, participants created rigid conditions forcing the AI to pause and discuss progress. For example, in P13's specification for conducting a literature review, she instructed the AI to \specquote{check in with the user to present this [coding] schema}, to \specquote{wait for user approval} of search criteria before initiating literature search, and to \specquote{schedule regular pings} to consult users on any ambiguities it runs into, etc (see Table \ref{tab:p13-spec} in the Appendix). Similarly, for their personalized note-taking application, P4 asked the AI to update a document every 24 hours that illustrated the current progress with findings, prototypes and screen recordings.

\end{enumerate}

Collectively, these AI specifications suggest a desire to create something akin to an \userquote{executable script,} which was intended to minimize ambiguity and tightly control the AI's workflow.\footnote{See Section \ref{sec:full-specs-examples} in the Appendix for a curated selection of full, non-confidential examples of these specifications.)} We further explore people's mental models behind these ``railway tracks'' in the following sections.



\subsection{Participants' mental models of humans vs. AI as observed in their specifications}\label{sec:findings-mental-models}

Our findings suggest that participants feel that current AI requires highly detailed, prescriptive instructions, compared to the more open-ended guidance given to people. In this section, we detail this contrast across four key dimensions (summarized in Table \ref{tab:mental-models}) that emerged from the data. The subsequent section will then contrast these findings with participants' vision of an ideal AI collaborator.

\subsubsection{\textbf{Reasoning and Judgment}}\label{sec:findings-mental-models-reasoning}

Participants perceived a fundamental difference in how AIs and humans exercise judgment and reason about instructions. They consistently characterized \textbf{AI as a ``literal executor''} that would follow the \textit{letter} of an instruction but not its \textit{spirit}. This mental model assumes that AI lacks the critical ability to \textit{infer unstated context}, \textit{question a flawed premise}, or \textit{strategically prioritize tasks}---skills participants considered the core ``value of a human'' collaborator. As P3 articulated, \userquote{Choosing what subset [of requirements] to build first is actually where a lot of the value of a human comes in [as opposed to the AI].}

The perceived inability to grasp the \userquote{bigger picture} (P4) was a notable concern. Participants feared that without a complete and unambiguous depiction of the project's goals and constraints, an AI could produce a \userquote{technically functional but practically unnecessary solution} (P6). For example, P10 worried that an AI simply told to \userquote{use a database} might default to a sophisticated solution like Postgres and create a \userquote{huge dev-ops nightmare.} He contrasted this with a human collaborator, who would be expected to understand the spirit of the goal and unstated project needs and select a more appropriate, easy to maintain option like Firebase.

To mitigate these risks, participants adopted the strategy of decomposing high-level goals into \textbf{highly prescriptive, step-by-step instructions}, oftentimes paired with explicit evaluation criteria. This approach was intended to make specifications \userquote{more executable} and a \userquote{closer description of what I want to make} (P2), thereby preventing the AI from taking what P5 called \userquote{the easiest possible route} and \userquote{prematurely call[ing] the task done.} For instance, P7 tasked an AI with providing research feedback by specifying a detailed workflow: generate keywords, find papers, produce a novelty score from 0-100, and explain its reasoning. Similarly, P3's specification for building a script writing assistant included a granular list of UI components (e.g., suggested titles, strapline, characters, story arc, themes) and precise commands for conversational turns, hoping that it would leave little room for the AI to deviate from his intended design.

In stark contrast, participants viewed a \textbf{human collaborator as an interpretive partner}. This partnership hinges on people's  capacity to reason about a collaborator's underlying intent, allowing one to navigate ambiguity, fill in missing requirements, and even question flawed assumptions in a project plan. Participants trusted their colleagues' ability to \userquote{read between the lines} (P9) to uncover a project's true, underlying goals. P10 described this as a human's essential role: to \userquote{reverse engineer the unspoken and probably unexplored... fractal of additional requirements.} While the latest generation of AI models may show traces of reasoning about user intent, participants' beliefs and prescriptive instructions discussed above suggest that they do not yet trust the AI in the same ways they trust colleagues.

Acting based on the assumption that a human does have this capability to infer intent and make informed choices, specifications written for them were therefore not final blueprints, but rather starting points of a collaborative discovery process. They were \textbf{intentionally conceptual and open-ended}, often framing high-level ``dream outcomes'' rather than concrete tasks. For example, P2's specification provided a ``big picture'' goal---\userquote{to design and develop a new user interface for large models that can mitigate people's overreliance on them}---while explicitly noting it was left \userquote{open-ended enough to leave room for pivoting.} Likewise, P8's spec described success through aspirational outcomes like \specquote{Convincing the relevant stakeholder/teams to invest resources} and \specquote{Teams feeling confident, excited, and energized to build towards the vision you help co-define.} Rather than offering prescriptive steps, participants embraced ambiguity, expecting a human collaborator to devise innovative ways to solve problems.

Furthermore, these specifications frequently included explicit invitations for the collaborator to exercise judgment and apply creative problem-solving skills. P14 directly solicited feedback in her specification, asking, \specquote{Are these above what you think as well?} and \specquote{What recent works from these venues are most similar?} Similarly, P11's specification scaffolded the development process in ``broad strokes,'' such as to \specquote{create sketches and an initial design specification... hold regular and iterative design sessions with the team to refine the design,} while leaving the minute details for the collaborator to determine. This demonstrates a trust in a human's ability to critically reflect, contribute expertise, and shape the project's direction through collaborative dialogue.

\subsubsection{\textbf{Agency and Drive}}\label{sec:findings-mental-models-agency}

Beyond the ability to reason, participants perceived a fundamental difference in the nature of AI versus human agency. They characterized the \textbf{AI as having a drive to ``blindly execute''}--a powerful but unchecked executional momentum that stems from its training to follow instructions. Participants described this as the AI's tendency to \userquote{steamroll right ahead} (P8) until it produces a final artifact. This behavior was not necessarily seen as an inherent inability to assess its own uncertainty, but rather as the default mode for an instruction-tuned model to always \userquote{give it its best shot} (P16) without pausing. Participants assumed that when faced with ambiguity or a false premise, an AI would not stop to reflect but would instead \userquote{pick one [direction] and go all in on it} (P3), risking a string of unproductive work. This fear of uncontrolled momentum, where \userquote{if it goes wrong it's a lot harder to debug} (P4), led participants to feel responsible for explicitly designing ways to \userquote{have the system stop... [and] tell you when it's uncertain} and ask for help.\looseness=-1

To counteract this ``steamrolling'' tendency, participants' primary strategy was to \textbf{impose external control} through their specifications. Thus, these documents became `\textbf{`command-and-control'' scripts filled with explicit ``tripwires''}---predefined conditions for pausing work---designed to regulate the AI's otherwise continuous workflow. For instance, participants defined strict and frequent check-in criteria and cadences to enforce these pauses: P8's specification included a mandatory daily user-briefing covering three key areas: \specquote{the most interesting things you've discovered/learned,} \specquote{what you're going to be working on tomorrow,} and whether its \specquote{thinking on anything has changed with the things uncovered today.} Similarly, after initially drafting a list of to-dos for the AI to complete, P13 felt \userquote{intimidated that... giving the AI this list of instructions somehow has some sort of like, finality to it,} which prompted her to add multiple explicit checkpoints where the AI had to seek her explicit feedback and approval before proceeding to the next action item. 

The assumption that AI required explicit, detailed control ironically led participants to explicitly indicate when they wanted the model to break \textit{out} of strict instruction following to \textit{more freely explore}. For example, the same specification from P8 that established rigid daily check-ins also stated, \specquote{otherwise feel free to explore as broadly and widely as you see fit.} Such explicit permissions for exploration were notably absent from specifications for humans, for whom creative latitude was considered the default, not an exception requiring a special instruction.

These specifications also included mechanisms to \textbf{manage uncertainty}. Some participants used direct encouragement, such as P8's instruction: \specquote{If you are ever uncertain... please reach out to me without waiting.} He hoped this would create opportunities for him to provide timely \userquote{feedback and steering} to make sure the effort that the AI spends is \userquote{worthwhile.} Others scaffolded a process of reflection; for example, P4 required the AI to maintain a \specquote{living doc} with links to work-in-progress prototypes and demos and report on \specquote{what things are being learned} so he could \userquote{re-orient [the AI] to the most interesting things} daily. It is worth noting that, beyond preventing unproductive work, participants also hoped that these check-ins would help them \userquote{slow down} the AI so they could \userquote{keep up with} its rapid progress (P6, P7, P10, P14).

Conversely, a \textbf{human collaborator was seen as having ``motivated agency.''} Participants suggested that a human's drive is usually internal, shaped by a need for inspiration, sense of ownership in the work, as well as professional rapport. As P14 articulated, the dynamic with a human is fundamentally relational: \userquote{If I have a real human on my team, that's a relationship now... The biggest thing [with humans] is relationship management. All [other] tasks are kind of... auxiliary.} From this perspective, the user's priority shifts from controlling momentum to aligning a collaborator's intrinsic motivation with the project's goals.

This mental model led participants to write specifications that were designed to be \textbf{motivating and enticing}. P3 noted that for a human partner, \userquote{you want the project to be open-ended enough to feel ambitious,} with P2 additionally suggesting that it could help \userquote{attract people who have different interests.} P8 emphasized that \userquote{motivation is kind of most important, everything flows from this,} using his specification to set an inspiring vision (\specquote{Welcome to our future-focused product team! We are a team of creatives who look 2-3 years into the future to find new opportunities...}, \specquote{Why now? Generative capabilities are improving exponentially... We need to move fast to stay ahead of the curve...}, etc.). In addition, the language in human-facing specs was also \textbf{supportive}, often using collaborative, question-based framing like \specquote{would you be interested in...?} and \specquote{what do you need from me to unblock you here?} (P14) to empower human agency and trust that a motivated collaborator would self-propel and regulate their own progress.

\subsubsection{\textbf{Knowledge and Expertise}}\label{sec:findings-mental-models-knowledge}

While participants assumed both AI and human collaborators possessed prior knowledge and expertise, their mental models for harnessing that knowledge appeared to be fundamentally different between the two. The \textbf{AI was perceived as having a vast but ungrounded knowledge base}. Its immense store of information from broad internet training was often seen not as a reliable asset, but as a latent and risky resource. Participants tended to distrust the AI's ability to correctly orient its knowledge for a specific task, fearing it might misinterpret key concepts or introduce bias. For example, P13 was concerned her specific research context around disabilities could be mishandled by the AI:

\begin{quote}
    \userquote{I was definitely assuming that the AI knew less about the specific area... I don't know where it's coming from in regard to disability representation... I know that oftentimes the AI responses are not great.}
\end{quote}

This led to a belief that the \textbf{AI's knowledge must be actively managed}. As P13 concluded, \userquote{I felt this need to... curate a specific set of background materials for the AI, to somewhat control what its initial grounding on this topic would be.}

This need to control the AI's grounding manifested in several distinct ways within the specifications. Some participants prescribed \textbf{explicit sources} to ``fence in'' the AI's knowledge, as P8 did by listing specific government websites (e.g., \specquote{[city] building department} and \specquote{planning department}), building codes, and social media forums (e.g., \specquote{Reddit threads}) that the AI must refer to while researching for a major home renovation project. Others, dissatisfied with default AI communication styles that tend to be \userquote{too verbose} (P7) or didn't \userquote{feel [...] human enough} (P2), supplied \textbf{preferred tones and styles}, such as \specquote{the language [of the output] should be approachable yet still highly credible} (P9). Finally, this control extended to dictating \textbf{entire professional workflows} on some occasions. For example, P13 provided a granular, step-by-step guide for conducting an academic literature review, from developing search criteria and a coding schema to synthesizing findings. This prescriptive guidance reflects a strategy to constrain the vast number of ways a task could be accomplished, ensuring the AI followed established and best practices or the user's known, preferred methodologies.

In contrast, a \textbf{human collaborator was viewed as a trusted, specialized expert} who is assumed to share significant \textbf{common ground} with the user. This common ground has been established through a prior vetting and alignment process, like hiring or team formation, which create a shared foundation of expertise and context before a project even begins (P11). Notably, this entire context-building journey is absent for the AI, placing the full burden of creating shared context onto the user, as we discussed above. The goal of a specification for a human is therefore to orient their existing, vetted expertise, not to construct their foundational understanding from scratch or to fence off problematic knowledge. P12 captured this core assumption of competence of a human expert:

\begin{quote}
    \userquote{I imagine if I hired a human... I hired them because they know where to go. And, for example, if they don't know to go to the planning department, then like, I shouldn't hire them in the first place...}
\end{quote}

With this foundation of common ground in place, specifications for human collaborators could be intentionally \textbf{lean documents} that deferred to their expertise, omitting details and norms they are expected to know and follow. P12 noted that for a human, he would \userquote{just keep the first two paragraphs [about project motivations] and get rid of everything afterwards,} including the aforementioned long list of reference sources for an AI. Some further reflected that they would not tell a competent peer \userquote{how to do their job} (P9) or \userquote{how to communicate with me} (P8), assuming a level of professional skill that they had to explicitly define for the AI. In sum, participants believed that the interaction with a human expert should be based on leveraging a trusted asset, not controlling an unpredictable one.

\subsubsection{\textbf{Capacity, Availability, and Costs}}\label{sec:findings-mental-models-cost}

Finally, a significant dichotomy emerged in how participants perceived the capacity and cost of AI versus human attention. The \textbf{AI was viewed as an infinitely expandable resource}, and perceived as having limitless capacity and low-cost scalability. In contrast, \textbf{human collaborators were viewed through the lens of a scarce attention economy}, where their cognitive bandwidth was a precious and expensive resource that must be carefully managed. 

As mentioned, participants operated as if the AI had near-limitless capacity for ingesting information, which drove a key difference in how they prepared instructions and contexts. For the AI, they often created exhaustive \userquote{context dumps} (P14), providing vast amounts of unfiltered information. While aware that this approach could be wasteful—a human would be expected to filter and selectively read such material—participants assumed the AI could effortlessly ingest an entire corpus without being \userquote{overwhelmed or insulted} (P11). This was evident in the pointers to massive data sources and highly detailed, step-by-step guides for AI discussed previously, and as P6 put it, \userquote{I'd be like, here's every file I've ever written, study this whole code base...} In contrast, when addressing humans, participants took on the cognitive burden of curating their \userquote{rambling thoughts} (P3) into a concise and coherent narrative. They did this to avoid being overwhelming or impolite, keenly aware that, as P14 stated, \userquote{With humans, human attention is very expensive... I would not give them all 5,000 files at once...}

Beyond AI's capacity to ingest information, participants leveraged the AI's perceived low cost to imagine fundamentally new, expandable workflows that would be infeasible with human collaborators. P9 candidly articulated this power-user mindset as her \userquote{flying car assumption that... these agents are free and their time is free.} This led her to design a competitive actor-critic model in her specification, with multiple AI duos working in parallel, each competing to produce the output the user would favor at the next time step—a strategy that treats AI agents as abundant and expendable resources. Similarly, P7 asked an AI to \specquote{do a full scanning of the literature... daily} across multiple research domains, a scale and frequency previously unimaginable for a human collaborator. P4 summarized how this mental model could affect project scoping entirely:

\begin{quote}
    \userquote{When you have people on a team, [...] their time is the most important thing. So if someone is excited about four projects, you need to cut it down to one... With LLMs and AI, I feel like you don't have to do that. They can do all 10 [projects] and then you can look at the outcomes.}
\end{quote}

However, it is important to note that the mental model of ``free AI'' was likely shaped by participants' context within a large tech corporation where the direct cost of computation is largely abstracted away; \savnew{this mindset likely does not reflect current non-``lead user'' usage of these models, where they are personally responsible for API costs.}
Nevertheless, as we will discuss in Section \ref{sec:discussion}, this perspective offers a valuable glimpse into a future where abundant and accessible AI resources could fundamentally reshape how complex problems are approached and solved across different lines of work.

\subsection{The Ideal Communication Model: Human-like Partnership with Tool-like Efficiency}\label{sec:findings-ideal-ai}

While participants' specifications were heavily shaped by their current mental models of AI, their vision for an ideal AI collaborator differed in very notable ways: they sought a partner that possesses the reasoning and proactive qualities of a trusted human expert, but can be communicated with using the directness and efficiency of a machine.

\subsubsection{\textbf{Proactive Thought Partner}}\label{sec:findings-ideal-ai-proactive}

In contrast to current AI interactions, participants wanted the AI to be more than a ``literal executor''---they wanted it to take on the role of a \textbf{proactive thought partner}, a collaborator that contributes from the early, ambiguous stages of planning through to the critical moments of execution. The desire for a proactive thought partner was partially motivated by participants acknowledging that people often \userquote{don't necessarily have a clear picture of what [they] want} (P9) at the start of a project. Thus, they imagined an AI that wouldn't just receive a specification but would actively \userquote{guide them through the specification authoring process, identifying and resolving ambiguities} with them (P4), effectively co-creating the plan rather than just executing it. 

Additionally, participants wanted AI to engage more critically during execution, reasoning about users' underlying intent, and \userquote{push[ing] back against flawed assumptions} (P3) when needed. For instance, P10 wished that, instead of defaulting to a technically correct but contextually poor and overly-complicated solution like Postgres, an AI would proactively question that initial plan and help uncover hidden requirements that are yet to be discovered: \userquote{...[I wish] the AI has the creativity to interrogate you and say like, `let me stop here for a second, …what is it you really want?'}

Furthermore, participants hoped that an AI could maintain a persistent memory and evolve its own understanding of the project over time, developing a mental model that allows it to \userquote{make unexpected connections} (P7) and offer novel insights that \userquote{spark new thinking} (P8). Rather than being a stateless tool that requires re-contextualization at the start of each conversation, participants envisioned an AI capable of supporting long-horizon projects where accumulated context and interaction would far exceed the current context window limitations of today's AI models.

\subsubsection{\textbf{Iterative Dialogue with Tight Feedback Loop}}\label{sec:findings-ideal-ai-iterative}

Secondly, participants advocated for an agile model of interaction with a \textbf{tight feedback loop}, enabling them to stay in the \userquote{driver's seat} (P9) through real-time steering and control. The current norm of writing an upfront specification often stood in direct tension with this desire, leading to anxiety about having to cede control to an AI that, by default, would ``steamroll ahead'' for extended periods of time. This concern about the AI's tendency to ``blindly execute'' drove participants to prefer to guide the AI through continuous, low-latency exchanges and frequent course corrections. P8 illustrated this anxiety with a ``Mars rover'' analogy, capturing the high stakes of long-latency commands: \userquote{It's like sending instructions to a Mars rover... you push left and it's gonna [take] 18 minutes... Oh man, what if I pushed the wrong button...}

Similarly, P11 expressed a desire for an agile approach to counter the sense of \userquote{finality} she associated with the large, upfront specifications she felt compelled to write for today's AI systems. Rather than relying on a single, high-uncertainty \userquote{waterfall} instruction set, she envisioned an iterative exchange, where plans could be safely adapted in response to both the AI's performance as well as emerging results, ensuring both human and AI efforts remain productive and worthwhile throughout the process.

\subsubsection{\textbf{Direct and Unburdened Communication}}\label{sec:findings-ideal-ai-hybrid}

Finally, while participants desired human-like reasoning from an AI, they explicitly wanted to \textbf{avoid the social overhead and attention constraints} that come with human interaction. Their ideal communication style leverages the AI's perceived limitless capacity and generosity, allowing for a directness that would be inappropriate with a human. They admitted that they already practice this directness (e.g., through large ``context dumps'' with current AI systems), and were hoping to continue doing so without the need to explicitly and frequently provide motivations or to express gratitude to an AI. P11 articulated this clearly, expressing a desire to be \userquote{a little more direct} with an AI in a way that \userquote{if I did that a lot with a human, it might sound like I'm badgering them or not believing them.} This highlights a unique communication dynamic where AI is seen as a tool for unburdened interaction, free from the social nuances required in human exchanges.

\subsection{\mxl{User Requirements} for Long-horizon AI}\label{sec:findings-user-req-for-long-horizon-ai}
\savvas{Collectively, the divergence between the rigid specifications participants currently write and the collaborative partnership they want to have with AI points to three \mxl{user requirements} for long-horizon AI:} 

\begin{enumerate}[leftmargin=0.2in]
\item \savvas{\textbf{Outcome alignment}. Participants mentioned they often \userquote{don't necessarily have a clear picture of what [they] want} (P9), and therefore want AI to help \userquote{guide them through the specification authoring process} (P4). This points to a need to help users discover the kind of outcome they want and align with AI on \textbf{what} it will create.}

\item \savvas{\textbf{Feasibility assessment}. Participants struggle with and feel the need to provide a complete set of the project's requirements to prevent the AI from producing a \userquote{technically functional but practically unnecessary solution} (P6). Therefore, another \mxl{requirement} is to help users verify \textbf{how} the AI intends to execute the task, including \mxl{its plans and processes, which tools it will use, the resources it needs access to, and crucially, how it plans to navigate roadblocks and recover from errors.}}

\item \savvas{\textbf{Monitoring and checking in on AI}. Participants wrote extensive check-in conditions to \userquote{slow down} AI and keep up with its \userquote{rapid progress} (P6, P7, P10, P11), pointing to a need to help users define these conditions in which AI should reach out to them in its long-running work.} 

\end{enumerate}

\section{Discussion}\label{sec:discussion}
This research focuses on the problem of using AI for long-horizon problems, problems that 1) an AI works on for an extended period of time on its own, 2) require communicating a significant amount of information to the AI (e.g., specifying the desired outcome, expectations for how to perform the work), and 3) often involve a fair amount of ambiguity (i.e., the desired outcome may not be completely known at the onset).
\mxl{In this section, we explore design implications stemming from the three user requirements identified in section \ref{sec:findings-user-req-for-long-horizon-ai} by proposing that future AI systems support aligning on outcomes through AI-generated rough drafts, verifying feasibility via end-to-end ``test runs,'' and monitoring execution through intelligent check-in criteria.}
\savnew{These implications are design hypotheses grounded in our empirical findings. They have not yet been prototyped or evaluated, and therefore, should be understood as directions for future system design and research.}

To make this discussion more concrete, we consider these design implications for an AI research assistant \mxl{(though we believe that these principles can equally apply to many other long-horizon scenarios, such as, coding, design, or administrative projects). In this particular scenario, the system} assists users in conducting open-ended information gathering and synthesis tasks. These workflows typically start with a simple user request (e.g., ``Write a literature review on the similarities and differences between orchestration of crowdworkers and LLM-powered AI agents in large systems''), and engage one or multiple AI agents that perform tasks such as planning, content search and filtering, and information synthesis. Systems like these provide an ideal context for discussing design implications, as they work for many minutes \mxl{(or hours)} at a time on more complex and ambiguous problems, with significant uncertainty in outcomes due to the need to search the web, deliberate between sources, and organize large amounts of information into a single, coherent whole \cite{openai_introducing_2025,google_deepmind_gemini_2025,perplexity_introducing_2025}.



\subsection{Aligning on the Outcome with AI Rough Drafts}

To help users iterate with AI and discover \textbf{what} they want, we propose that AI produce rough drafts of the final artifact(s).
The kind of rough draft it produces will depend on the nature of the ambiguity it is aiming to clarify with the user.
For example, the AI might produce a \textit{sketch of the full solution}, such as a bulleted list of themes to verify the literature review's scope. Alternatively, it might offer \textit{a sample, or high-fidelity excerpt} of the final output, such as a single, fully written paragraph to demonstrate the intended depth and tone.
With these drafts, users can recognize what they want and provide feedback, by pruning a theme in the outline, or adjusting the prose in the paragraph excerpt. This capability extends prior work on co-creating plans for AI \cite{feng_cocoa_2025}, which focused on decomposing problems into steps to provide transparency into how the agent works.
Here, the AI goes beyond reasoning about the process to explicitly reason about the intermediate artifacts, determining which type of draft will best serve to clarify the user’s goals. \mxl{Crucially, these drafts allow users to verify that the AI is oriented toward the ``compass heading'' that they had in mind before it expends resources on the full task.}

As the drafts become more involved, another challenge is reducing the \mxl{\textbf{cognitive load}} of providing feedback on them.
One path to explore is to leverage recent models' ability to generate user interfaces \cite{wu_uicoder_2024,google_a2ui_2025}, so the AI can present the draft as a dynamic and interactive artifact, instead of a static document.
The AI could reason about the interactions or tools that might be added to the draft to make providing feedback easier.
For example, for the literature review, instead of providing a static, long list of papers for the user to peruse, the AI might determine it's more engaging to give them the ability to delete or ``star'' potential papers to indicate their importance to a theme.
There might also be suggested actions within the rough draft for the user to directly manipulate the draft, like a ``find more papers'' button to fill out a theme a bit more.
And finally, the AI might preemptively leave comments in the rough draft, to highlight its uncertainties and proactively guide user feedback. These comments might also serve to highlight potential alternatives and prevent fixation on the AI’s initial output \cite{suh_luminate_2024}.
By reasoning about and producing rough drafts, AI functions more like the hybrid-collaborator participants wanted, using its processing power to complete a substantial amount of work and putting the user in the driver's seat to make key decisions.

\subsection{Verifying the Task's Feasibility with Test Runs}
To help users assess the feasibility of the project and see \textbf{how} the final artifact might be created, we propose that AI complete a ``test run'', where it works through a miniature, end-to-end version of the project.
For example, the AI can first complete a pared down version of the literature review, where it goes through all the steps (e.g., collecting papers, synthesizing themes, producing outlines), but with only a few papers.
By seeing the entire process end-to-end, the user can get a concrete glimpse into the kinds of barriers AI might encounter and the decisions it will make, before it spends a significant amount of time and resources on the full project.
\mxl{This approach shifts the user's cognitive burden from \textit{recalling} every ``unknown unknown'' and edge case in a vacuum while authoring the specification to \textit{recognizing} issues in the AI's workflow and the concrete artifact it produced \cite{liu_reuse_2021}.}

For instance, while searching for papers, the AI might realize it cannot access a useful paper repository (e.g., the ACM Digital Library), resorting to accessing all of its papers from ArXiv. 
Having seen this, the user can update their instructions with a list of preferred, easily accessible repositories for the agent to use alongside ArXiv.
Later on in the test run, the user might notice the AI skipped the step of producing themes prior to creating an outline for the paper and update their instructions to include this step.
Overall, these test-runs can help users anticipate the challenges the AI might face and further specify their project so that it executes more smoothly.

\subsection{Supporting User Check-ins to ``Slow Down'' AI}\label{sec:check-ins}

To help users monitor and check-in on AI, we propose that future systems reason about and help establish check-in criteria.
These systems can suggest to the user different scenarios in which it might check-in, including when the AI (1) achieves a significant milestone in the project, like producing an outline of the literature review, as well as when it (2) hits an edge case or ambiguity (e.g. the AI encounters very relevant pre-print papers and is not sure whether to include them, since they have not been published).
Prior to starting a project, the AI might reason about its potential milestones and ambiguities, and converse with the user to determine which of these it should check in with them on.
At the same time, however, as evidenced by our findings in section \ref{sec:findings-ideal-ai-hybrid}, these check-ins should not occur too frequently and should be low-friction, as users wanted high control, without the social overhead of constant management.

In addition to establishing these criteria, the AI might also leverage its processing power and reasoning capabilities to help users make informed decisions during a check-in.
For the pre-print ambiguity, instead of waiting for the user's response, the AI might reason about work it could do to help the user determine if they should include these papers or not.
For example, it might produce two potential narratives of the literature review, one that includes these pre-print papers and another that does not.
By helping to establish check-in criteria for a variety of scenarios and proactively complete work to make these check-ins more effective for the user, AI can function more like the hybrid-collaborator participants desired.

\section{Limitations \& Future Work}\label{sec:limitations}

Our findings represent a snapshot of users' mental models at a specific moment in the rapid evolution of AI. As AI capabilities advance, we anticipate that some of the challenges identified---such as the need for highly prescriptive, step-by-step instructions---may subside as systems become better at inferring intent. However, we expect some challenges to remain, even as systems continue to improve. For example, with any collaborator, one needs to communicate goals, ensure alignment, and establish ways of working. This study shows how current AI meets (or doesn't meet) these fundamental needs, thus serving as a valuable baseline against which future systems can be compared.

This work primarily focused on analyzing the ``first mile'' of collaboration: the initial specification of a long-running problem. 
We made this methodological choice because we observed that, for some use cases, prompts have evolved to become highly detailed specifications.  
While we did not observe the full, iterative execution of the task, isolating this specification phase allowed us to generate a rich, detailed understanding of users' mental models and processes for this important phase.
\mxlnew{
This also means we did not capture how specifications evolve when projects encounter shifting requirements, scope changes, or execution surprises—and some participants hinted that such revision would be necessary.  
The rigid, front-loaded ``railway tracks'' approach may be especially brittle in these scenarios; human delegation naturally absorbs midstream shifts through ongoing dialogue, a dynamic our proposed ``intelligent check-ins'' (section \ref{sec:check-ins}) are designed to support.
}

In addition, our findings are based on a small sample of technical experts (designers, researchers, and software engineers), which could limit the generalizability of our results.
Our primary goal was to generate deep qualitative evidence and conceptual insights into the emergent phenomenon of delegating complex work to AI, rather than to make broad population claims.
Furthermore, we argue that this specific group can be viewed as ``lead users'' \cite{von_hippel_lead_1986}, who are at the forefront of encountering the opportunities and challenges of advanced AI systems.
\mxlnew{
Their current practices foreshadow the challenges a broader user base will face. 
That said, we expect the four mental model dimensions to generalize more readily than the specific delegation strategies we observed. 
The prescriptive ``railway tracks'' style may be amplified by our participants' technical training and professional habit of writing detailed specifications; non-technical users facing the same trust gap might compensate differently—e.g., by avoiding delegation entirely or providing informal instructions rather than structured specs. 
Additionally, recruiting from a single institution with an AI-forward culture may have normalized certain attitudes not shared across disciplines, and we did not control for participants' prior track records with AI—both successes and failures—which likely shape delegation behavior.
Nonetheless, studying these lead users provides an early window into the next frontier of human-AI interaction and a foundation for future work with non-experts and professionals in other domains such as law, medicine, and the creative arts.
}

\savnew{While the compass and railway tracks framework likely reflects the predominant delegation pattern across a wide range of long-horizon tasks, future work should examine the limits and edge cases of this framework more carefully, as it may not hold across \textit{all} task types. 
For instance, even with human collaborators, there are domains, such as clinical protocols, legal compliance, or safety-critical engineering where prescriptive ``railway track'' instructions remain essential regardless of the collaborator's expertise, precisely because deviations carry high stakes.
Conversely, for AI handling highly verifiable, low-ambiguity tasks, like reformatting a large dataset or running a common statistical analysis, a ``compass''-style set of instructions may be sufficient, since errors are immediately detectable and easily corrected.
Investigating where these boundary conditions lie, and what task properties (e.g., reversibility, verifiability, stakes, and ambiguity) govern them, would provide further insight into how humans delegate tasks and offer more targeted design guidance for future human-AI collaboration tools.}

\mxlnew{
Finally, looking beyond this initial specification phase, we hypothesize that successful collaboration might follow a trajectory from ``railway tracks'' to ``compass'': as trust is established through rough drafts, successful test runs, and check-ins, users may gradually relax their constraints, eventually granting AI the open-ended agency they currently reserve for humans.  
This would mirror how human delegation evolves in organizations, where new hires receive detailed, step-by-step instructions, but as they build a track record and shared context with their manager, delegation shifts toward high-level goals and greater autonomy. 
Longitudinal studies tracking the same users over sustained AI collaboration could reveal whether this trajectory holds---whether the prescriptive ``railway tracks'' approach is a stable preference or an early-stage strategy that softens with calibrated trust.
Our three design implications---rough drafts, test runs, and intelligent check-ins---may serve as precisely the mechanisms through which such trust is built.
}

\section{Conclusion}\label{sec:conclusion}

\mxl{Our study reveals a fundamental divergence in how people currently communicate complex, long-horizon work to AI versus human colleagues: while users provide human colleagues with a ``compass'' of high-level intent that empowers them to navigate ambiguity, they constrain AI to ``railway tracks'' of rigid, exhaustive instructions to prevent deviation. This approach stems from a mental model of current AI as a powerful but brittle executor that struggles to infer intent, prioritize tasks, or self-correct its course.
Interestingly, users do not simply want a long-running AI that mimics a human. They instead envision an ideal ``hybrid collaborator'' that blends the critical agency of a human expert to question premises and clarify goals with the machine's scalability and freedom from social overhead, allowing for direct, unburdened communication. 
To bridge this gap, we highlight the value in evolving current human-AI interfaces to accommodate the novel needs of long-horizon AI, such as aligning on outcomes through generated rough drafts, verifying feasibility via end-to-end test runs, and monitoring execution through intelligent check-ins---ultimately transforming AI from a passive instruction-follower into a reliable partner for ambiguous, long-term problems.}

\begin{acks}
We gratefully thank Alex Bäuerle, Omar Benjelloun, Andy Coenen, Noah Fiedel, Tesh Goyal, Tianchang He, Ellen Jiang, Sherry Moore, Merrie Morris, Mahima Pushkarna, Fernanda Viegas, Zi Wang, Martin Wattenberg, and James Wexler for their helpful comments and feedback. 
We also appreciate the valuable input from our study participants.
\end{acks}



\bibliographystyle{ACM-Reference-Format}
\bibliography{chi-2026-spec-references}

\onecolumn 

\appendix

\section{Selected Full Specification Examples}\label{sec:full-specs-examples}
Note on confidentiality: \textit{Participants were encouraged to write specifications for active, real-world projects relevant to their current roles and work. As a result, several submissions contained confidential internal information. The examples presented in this section represent a subset of the collected data that could be publicly shared without redaction.}

\definecolor{subtleblue}{RGB}{10, 80, 140}
\newcolumntype{P}[1]{>{\raggedright\arraybackslash}p{#1}}

\def\arraystretch{1.1} 

\begin{table*}[h]
\centering
\caption{Specification examples from P13, particularly illustrating the various ``check-in policies and conditions'' that participants provided in their AI specifications.}
\label{tab:p13-spec}
\small 
\resizebox{1\textwidth}{!}{%
\begin{tabular}{P{86mm} | P{78mm}}
\toprule
\multicolumn{2}{c}{\textbf{P3: Design and build a script writing assistant}} \\ 
\midrule

\textbf{Specification for Human} & 
\textbf{Specification for AI}  (\textit{with ``prescriptive outputs'' highlighted in blue}) \\
\midrule

Writers often start with a very broad idea of what they want to write about and the themes they want to explore. Script-buddy is a Web-based companion that will help writers develop their ideas. The core interaction is a two-way conversation with an AI, and it should ask clarifying questions and create various artifacts that will flesh out the script plan off the back of the conversation.\par
\vspace{1mm}
This is an exploratory project so use this as inspiration only.\par
\vspace{1mm}
\textbf{Requirements}
\begin{itemize}[leftmargin=*, nosep]
    \item The interaction is voice-to-voice
    \item Visual elements and artifacts are created and updated (e.g. character list and story arc) during the conversation
\end{itemize}
\vspace{1mm}
\textbf{Conversation Structure:}
\begin{itemize}[leftmargin=*, nosep]
    \item AI should ask clarifying questions to draw out more detailed descriptions.
    \item AI should have access to tvtropes.org to be able to suggest well-known tropes that the user idea might be close to.
\end{itemize}
\vspace{1mm}
\textbf{Architecture:} I'd suggest an Angular app talking to an LLM/AI API. \textit{[potential diagram they are free to deviate from]} 
&
\textit{[removed motivation]}\par
\vspace{1mm}
You are an expert engineer and your task is to create an angular application to allow a writer to co-develop a new idea for a film.\par
\vspace{1mm}
{\color{subtleblue}
\textbf{This is an example conversation:}
\begin{itemize}[leftmargin=*, nosep]
    \item AI: \textit{[to be filled in]}
    \item User: \textit{[to be filled in]}
    \item AI: \textit{[to be filled in]}
    \item User: \textit{[to be filled in]}
\end{itemize}
\vspace{1mm}
\textbf{This conversation should generate the following artifacts:}
\begin{itemize}[leftmargin=*, nosep]
    \item Character List and character summaries. Examples: \textit{[to be filled in]}
    \item Overall story arc: this should be based on the standard 5-act story-structure. \textit{[Define: 5-act story structure.]}
    \item Themes: Examples: \textit{[to be filled in]}.
    \item Title suggestions.
    \item Strapline.
\end{itemize}}
\vspace{1mm}
\textbf{UI Sketch:} \textit{[to be filled in]} \\
\bottomrule
\end{tabular}%
}
\caption{Specification examples from P3, particularly illustrating the ``prescriptive outputs'' that participants provided in their AI specifications.}
\label{tab:p3-spec}
\end{table*}

\begin{table*}[h]
\centering
\small 
\resizebox{1\textwidth}{!}{%
\begin{tabular}{P{60mm} | P{100mm}}
\toprule
\multicolumn{2}{c}{\textbf{P12: Research and plan for a major home renovation}} \\ 
\midrule

\textbf{Specification for Human} & 
\textbf{Specification for AI}  (\textit{with ``explicit information sources'' highlighted in blue}) \\
\midrule

I need your help going deep and assembling a literature review about all the San Francisco building codes, rules, regulations, possible pitfalls, and other things to know when remodeling a single family home in San Francisco.\par
\vspace{1mm}
The ultimate goal is to help me remodel my home in San Francisco and navigate the (from what I’ve heard to be) incredibly difficult planning and building process in SF. I’m trying to educate myself, build confidence, and avoid common mistakes to help make this process more efficient. 
&
\textit{[everything from human spec]}\par
\vspace{1mm}
I want you to go as broad as possible and review everything from:
{\color{subtleblue}
\begin{itemize}[leftmargin=*, nosep]
    \item San Francisco Building Department
    \item San Francisco Planning Department
    \item Recent single family home renovations that have been done in the Outer Richmond neighborhood
    \item The history of building in San Francisco and how the building codes have changed over the years and why
    \item Any social media posts with first hand experience from others that have gone through this process recently
    \item Anything else that you think I will need to get a full picture of this topic
\end{itemize}
} \\
\bottomrule
\end{tabular}%
}
\caption{Specification examples from P12, particularly illustrating the ``explicit information sources'' that participants provided in their AI specifications.}
\label{tab:p12-spec}
\end{table*}

\begin{table*}[h]
\centering
\small 
\resizebox{1\textwidth}{!}{%
\begin{tabular}{P{48mm} | P{112mm}}
\toprule
\multicolumn{2}{c}{\textbf{P13: Disability-related sociotechnical harms: a literature review}} \\ 
\midrule

\textbf{Specification for Human} & 
\textbf{Specification for AI} (\textit{with ``check-in policies and conditions'' highlighted in blue}) \\
\midrule

\textbf{Background}: As gen AI proliferates and offers more interactive experiences, researchers are identifying, evaluating, and mitigating sociotechnical harms that may be perpetrated by the AI, and/or by users employing AI tools to do work and make decisions. Researchers have not developed a taxonomy of sociotechnical harms related to disability. This gap means that we do not know if identified harms impact people with disabilities in the same way, and we don’t know if there are unique harms that disproportionately impact people with disabilities. Fortunately, researchers in many fields such as sociology, disability studies, and others, have documented sociotechnical harms both related to and beyond technology. Understanding these harms can help us to identify, evaluate, and mitigate such harms in AI systems. \par
Your goal is to bridge this gap by conducting a systematic literature review:  \par
\begin{itemize}[leftmargin=*, nosep]
    \item Start by searching for and analyzing literature that bridges disability studies and technology.
    \item You will then help build the framework (taxonomy) that categorizes these harms.
    \item You will help draft a paper that establishes a new standard for how AI researchers evaluate disability-related risks.
\end{itemize}
Feel free to set up regular 1:1s to discuss strategies, questions, ambiguities, interim results, or anything on your mind. 
&
\textbf{Background}: As gen AI proliferates and offers... \textit{[the same background paragraph from human spec]}\par
\textbf{Instructions (must follow strictly)}: \par
\begin{itemize}[leftmargin=*, nosep]
    \item Review the provided background information regarding disability and sociotechnical harms. Internalize this framing to ensure all subsequent outputs align with these specific project goals and theoretical groundings.
    \item Develop literature search criteria.
    \item Develop a schema for annotating potentially-relevant articles; {\color{subtleblue} check in with the user to present this schema and discuss how to categorize ``edge cases'' that may or may not be included before proceeding to the full literature review.}
    \item Refine the literature inclusion criteria based on user feedback, also ask the user for any systematic lit review method they used in the past for additional references; {\color{subtleblue} wait for user approval of the finalized criteria list before initiating the formal search.}
    \item Conduct the literature search with all of the inclusion criteria.
    \item Organize all found articles in a user designated storage location; create an annotated bibliography of the articles, capturing metadata and high-level summaries for each entry. {\color{subtleblue} Notify the user once the bibliography is ready for review.}
    \item Pick and read through a few articles and generate detailed summaries of these articles; based on these summaries, propose an initial data coding schema. {\color{subtleblue} Check-in with the user to review these summaries and the proposed schema before any formal data coding begins.}
    \item Apply the proposed coding schema to another small sample of articles while reminding the user to do the same. Compare these results with the user’s manual coding. {\color{subtleblue} Pause here to refine definitions and schema categories based on the user’s feedback and findings.}
    \item Apply the refined coding schema to the remaining articles. {\color{subtleblue} Schedule regular pings (e.g., every 5–10 articles) to present edge cases or ambiguous data points.}
    \item Organize the coded data into a structured "Findings" section, categorized by the inclusion criteria and the codes we developed through the refined code development process. {\color{subtleblue} Check-in with the user to ensure the findings emphasize novelty and correctly relate to prior work.}
    \item Iterate on the ``Findings'' section based on user feedback.
    \item Propose a list of the most significant or ``interesting'' themes to serve as the scaffold for the ``Discussion'' section. Wait for user confirmation on which themes to prioritize.
    \item Draft the "Discussion" section of the paper.
    \item Review and iterate on the full paper draft for improvements. {\color{subtleblue} Engage the user for feedback after each revision cycle.}
    \item Update the annotated bibliography by integrating specific findings and codes for every article reviewed. Generate a comprehensive appendix that includes all data points, not just those included in the paper. {\color{subtleblue} Notify the user once the final appendix is ready for review.}
\end{itemize} \\
\bottomrule
\end{tabular}%
}
\caption{Specification examples from P13, particularly illustrating the various ``check-in policies and conditions'' that participants provided in their AI specifications.}
\label{tab:p13-spec}
\end{table*}

\end{document}